\def\year{2020}\relax
\documentclass[letterpaper]{article} 
\usepackage{aaai20}  
\usepackage{times}  
\usepackage{helvet}  
\usepackage{courier}  
\usepackage{url}  
\usepackage{graphicx}  
\usepackage{array,multirow}
\usepackage{xcolor}
\usepackage{array}
\usepackage{booktabs}
\usepackage{enumitem,kantlipsum}
\usepackage[normalem]{ulem}

\newcommand{\rev}[1]{{\color{black}#1}}

\newcolumntype{L}[1]{>{\raggedright\let\newline\\\arraybackslash\hspace{0pt}}m{#1}}
\newcolumntype{C}[1]{>{\centering\let\newline\\\arraybackslash\hspace{0pt}}m{#1}}
\newcolumntype{R}[1]{>{\raggedleft\let\newline\\\arraybackslash\hspace{0pt}}m{#1}}
\begin{document}
%
\title{Unsupervised User Stance Detection on Twitter}
\author{
Kareem Darwish$^1$, Peter Stefanov$^2$, Micha\"el Aupetit$^1$, Preslav Nakov$^1$\\
$^1$ Qatar Computing Research Institute\\
Hamad bin Khalifa University, Doha, Qatar\\
$^2$ Faculty of Mathematics and Informatics, \\
Sofia University "St. Kliment Ohridski", Sofia, Bulgaria \\
kdarwish@hbku.edu.qa, p.stefanov@hotmail.com,  maupetit@hbku.edu.qa, pnakov@hbku.edu.qa\\
}

\maketitle
\begin{abstract}
We present a highly effective unsupervised framework for detecting the stance of prolific Twitter users with respect to controversial topics. In particular, we use dimensionality reduction to project users onto a low-dimensional space, followed by clustering, which allows us to find core users that are representative of the different stances.
Our framework has three major advantages over pre-existing methods, which are based on supervised or semi-supervised classification.  First, we do not require any prior labeling of users: instead, we create clusters, which are much easier to label manually afterwards, e.g.,~in a matter of seconds or minutes instead of hours. Second, there is no need for domain- or topic-level knowledge either to specify the relevant stances (labels) or to conduct the actual labeling. Third, our framework is robust in the face of data skewness, e.g.,~when some users or some stances have greater representation in the data. We experiment with different combinations of user similarity features, dataset sizes, dimensionality reduction methods, and clustering algorithms to ascertain the most effective and most computationally efficient combinations across three different datasets (in English and Turkish). \rev{We further verified our results on additional tweet sets covering six different controversial topics.} 
Our best combination in terms of effectiveness and efficiency uses retweeted accounts as features, UMAP for dimensionality reduction, and Mean Shift for clustering, and yields a small number of high-quality user clusters, typically just 2--3, with more than 98\% purity. The resulting user clusters can be used to train downstream classifiers. Moreover, our framework is robust to variations in the hyper-parameter values and also with respect to random initialization.  
\end{abstract}

\section{Introduction}

Stance detection is the task of identifying the position of a user with respect to a topic, an entity, or a claim \cite{mohammad2016semeval}, and it has broad applications in studying public opinion, political campaigning, and marketing. Stance detection is particularly interesting in the realm of social media, which offers the opportunity to identify the stance of very large numbers of users, potentially millions, on different issues. Most recent work on stance detection has focused on supervised or semi-supervised classification. 

\noindent In either case, some form of initial manual labeling of tens or hundreds of users is performed, followed by user-level supervised classification or label propagation based on the user accounts and the tweets that they retweet and/or the hashtags that they use \cite{magdy2016isisisnotislam,pennacchiotti2011democrats,wong2013quantifying}.

Retweets and hashtags can enable such classification as they capture homophily and social influence \cite{DellaPosta2015latte,magdy2016isisisnotislam}, both of which are phenomena that are readily apparent in social media. With homophily, similarly minded users are inclined to create social networks, and members of such networks exert social influence on one another, leading to more homogeneity within the groups. Thus, members of homophilous groups tend to share similar stances on various topics~\cite{garimella2017quantifying}. Moreover, the stances of users are generally stable, particularly over short time spans, e.g.,~over days or weeks. All this facilitates both supervised classification and semi-supervised approaches such as label propagation. Yet, existing methods are characterized by several drawbacks, which require an initial set of labeled examples, namely: (\emph{i})~manual labeling of users requires topic expertise in order to properly identify the underlying stances; (\emph{ii})~manual labeling also takes substantial amount of time, e.g.,~1--2 hours or more for 50--100 users; and (\emph{iii})~the distribution of stances in a sample of users to be labeled, e.g.,~the $n$ most active users or random users, might be skewed, which could adversely affect the classification performance, and fixing this might require non-trivial hyper-parameter tweaking or manual data balancing.

Here we aim at performing stance detection in a completely unsupervised manner \rev{to tag the most active users on a topic, which often express strong views}. Thus, we overcome the aforementioned shortcomings of supervised and semi-supervised methods. Specifically, we automatically detect homogeneous clusters, each containing a few hundred users or more, and then we let human analysts label each of these clusters based on the common characteristics of the users therein such as the most representative retweeted accounts or hashtags. This labeling of clusters is much cheaper than labeling individual users. The resulting user groups can be used directly, and they can also serve to train supervised classifiers or as seeds for semi-supervised methods such as label propagation.

\begin{figure}[tbh]
    \centering
    \includegraphics[width=.85\linewidth]{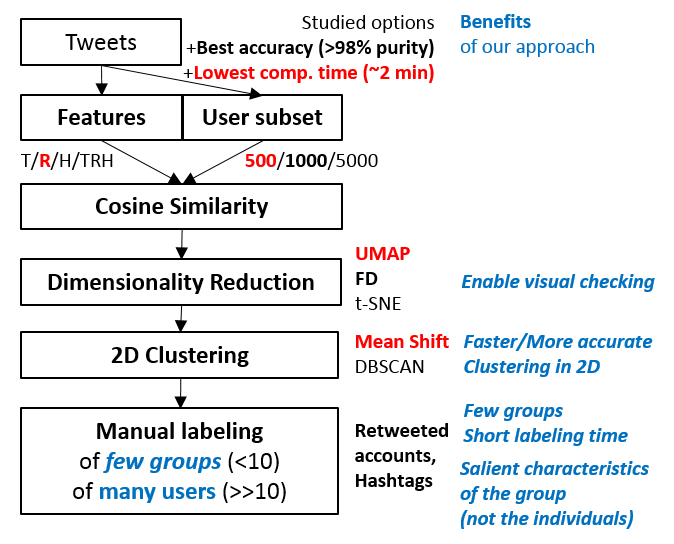}
    \caption{Overview of our stance detection pipeline, the options studied in this paper, and the benefits they offer. In bold font: best option in terms of accuracy. In bold red: the best option both in terms of accuracy and computing time.}
    \label{fig:pipelineoverview}
\end{figure}

\noindent Our method works as follows (see also Figure \ref{fig:pipelineoverview}): given a set of tweets on a particular topic, we project the most active users onto a two-dimensional space based on their similarity, and then we use peak detection/clustering to find core groups of similar users. Using dimensionality reduction has several desirable effects. First, in a lower dimensional space, good projection methods bring similar users closer together while pushing dissimilar users further apart. User visualization in two dimensions also allows an observer to ascertain how separable users with different stances are. 

Dimensionality reduction further facilitates downstream clustering, which is typically less effective and less efficient in high-dimensional spaces. Using our method, there is no need to manually specify the different stances a priori. Instead, these are discovered as part of clustering, and can be easily labeled in a matter of minutes at the cluster level, e.g.,~based on the most salient retweets or hashtags for a cluster.
Moreover, our framework overcomes the problem of class imbalance and the need for expertise about the topic. 

In our experiments, we compare different variants of our stance detection framework. In particular, we experiment with three different dimensionality reduction techniques, namely the Fruchterman-Reingold force-directed~(FD) graph drawing algorithm \cite{fruchterman1991graph}, t-Distributed Stochastic Neighbor Embeddings (t-SNE) \cite{maaten2008visualizing}, and Uniform Manifold Approximation and Projection (UMAP) algorithm \cite{mcinnes2018umap}.  For clustering, we compare DBSCAN \cite{ester1996density} and Mean Shift \cite{comaniciu2002mean}, both of which can capture arbitrarily shaped clusters.  We also experiment with different features such as \textit{retweeted users} and \textit{hashtags} as the basis for computing the similarity between users.  
The successful combinations use FD or UMAP for dimensionality reduction, Mean Shift for peak detection, and retweeted accounts to compute user similarity. We also explore robustness with respect to hyper-parameters and the required minimum number of tweets and users.

\noindent Overall, we experiment with different sampled subsets from three different tweet datasets with gold stance labels in different languages and covering various topics, and we show that we can identify a small number of user clusters (2-3 clusters) composed of hundreds of users on average with purity in excess of 98\%. We further verify our results on additional tweet datasets covering six different controversial topics.

Our contributions can be summarized as follows: \begin{itemize}
    \item We introduce a robust stance detection framework for automatically discovering core groups of users without the need for manual intervention, which enables subsequent manual bulk labelling of all users in a cluster at once.
    \item We overcome key shortcomings of existing supervised and semi-supervised classification methods such as the need for topic-informed manual labeling and for handling class imbalance and the presence of potential skews.
    \item We show that dimensionality reduction techniques such as FD and UMAP, followed by Mean Shift clustering, can effectively identify core groups of users with purity in excess of 98\%.
    \item We demonstrate the robustness of our method to changes in dimensionality reduction and clustering hyper-parameters as well as changes in tweet set size, kinds of features used to compute similarity, and minimum number of users, among others. In doing so, we ascertain the minimum requirements for effective stance detection.
    \item We elucidate the computational efficiency of different combinations of features, user sample size, dimensionality reduction, and clustering.
\end{itemize}

\section{Background}
\paragraph{Stance Classification: }
There has been a lot of recent research interest in stance detection with focus on inferring a person's or an article's position with respect to a topic/issue or political preferences in general~\cite{barbera2015birds,barbera2014understanding,borge2015content,cohen2013classifying,colleoni2014echo,conover2011predicting,fowler2011causality,himelboim2013birds,magdy2016isisisnotislam,magdy2016failedrevolutions,makazhanov2014predicting,mohtarami-etal-2018-automatic,mohtarami-etal-2019-contrastive,ACL2020:Topical:Stance,weber2013secular}.

\paragraph{Effective Features:} Several studies have looked at features that may help reveal the stance of users.  This includes \emph{textual features} such as the text of the tweets and hashtags, \emph{network interactions} such as retweeted accounts and mentions as well as follow relationships, and \emph{profile information} such as user description, name, and location \cite{borge2015content,magdy2016isisisnotislam,magdy2016failedrevolutions,weber2013secular}.  Using network interaction features, specifically retweeted accounts, was shown to yield better results compared to using content features \cite{magdy2016isisisnotislam}.  

\paragraph{User Classification:}
Most studies focused on supervised or semi-supervised methods, which require an initial seed set of labeled users. 
Label propagation was used to automatically tag users based on the \textit{accounts they follow} \cite{barbera2015birds} and \textit{retweets} \cite{borge2015content,weber2013secular}. Although it has very high precision (often above 95\%), it has three drawbacks: (\emph{i})~it tends to label users who are more extreme in their views, (\emph{ii})~careful manipulation of thresholds may be required, particularly when the initial tagged user set is imbalanced, and (\emph{iii})~post checks are needed.
Some of these issues can be observed in the \emph{Datasets} section below, where two of our test sets were constructed using label propagation. Our method overcomes the latter two drawbacks.

Supervised classification was used to assign stance labels, where classifiers were trained using a variety of features such as \textit{tweet text, hashtags, user profile information, retweeted accounts} or \textit{mentioned accounts}  \cite{magdy2016isisisnotislam,magdy2016failedrevolutions,pennacchiotti2011democrats}.  Such classification can label users with precision typically ranging between 70\% and 90\%. Rao et al.~\shortcite{rao2010classifying} used socio-linguistic features that include types of utterances, e.g.,~emoticons and abbreviations, and word $n$-grams to distinguish between Republicans and Democrats with more than 80\% accuracy.  Pennacchiotti and Popescu~\shortcite{pennacchiotti2011democrats} extended the work of Rao et al.~\shortcite{rao2010classifying} by introducing features based on profile information (screen name, profile description, followers, etc.), tweeting behavior, socio-linguistic features, network interactions, and sentiment. It has been shown that users tend to form so-called ``echo chambers'', where they engage with like-minded users~\cite{himelboim2013birds,magdy2016isisisnotislam},
and they also show persistent beliefs over time and tend to maintain their echo chambers, which reveal significant social influence \cite{borge2015content,magdy2016isisisnotislam,pennacchiotti2011machine}.
Duan et al.~\shortcite{duan2012graph} used the so-called ``collective classification'' techniques to jointly label the interconnected network of users using both their attributes and their relationships. Since there are implicit links between users on Twitter (e.g.,~they retweet the same accounts or use the same hashtags), collective classification is relevant here. Darwish et al.~\shortcite{darwish2017improved} extended this idea by employing a so-called user similarity space of lower dimensionality to improve supervised stance classification. There was a related SemEval-2016~\cite{mohammad2016semeval} task on stance detection, but it was at the tweet level, not user level.

\paragraph{Dimensionality Reduction and Clustering:} 
\rev{A potential unsupervised method for stance detection may involve user clustering.} Beyond the selection of relevant features for stance detection, a major challenge for clustering approaches is the number of features. Indeed, an expert may be willing to use as many meaningful input features as possible, expecting the machine to detect  automatically the relevant ones for the task at hand. This high-dimensional space is subject to the \emph{curse of dimensionality} \cite{VerleysenCurseDim2003}:
the search space for a solution grows exponentially with the increase in dimensionality as there are many more possible patterns than in lower-dimensional subspaces; and the computation time and the memory needed for clustering also grow. Moreover, it has been shown that as dimensionality increases, the distance from any point to the nearest data point approaches the distance to the furthest data point \cite{beyer1999when}. This is problematic for clustering techniques, which typically assume short within-cluster and large between-cluster distances. We conducted experiments that involved clustering directly in the high-dimensional feature space and all of them failed to produce meaningful clusters.  On the other hand, most clustering techniques are very efficient in low-dimensional spaces.

\rev{Another issue comes from the need for human experts to ascertain the validity of the clustering result beyond standard clustering statistics. For instance, an expert may want to verify that users belong to the core of separable groups such that they are good representatives of the groups and good candidate seeds for possible subsequent classification. 

Visualization has come as a natural way to support the experts using Dimensionality Reduction (DR) or   Multi-Dimensional Projection (MDP)  \cite{AupetitNonatoMDPsurvey2018}. 
Different pipelines combining Dimensionality Reduction and Clustering have been studied \cite{DRclusterPipelineVA_Wenskovitch2018} in Visual Analytics in order to support user decision, giving guidelines to select the best approach for a given application. As our primary goal is to support users to check cluster quality visually and label data based on cluster information, and given that clustering is more efficient in low dimensionality, we decided to first reduce data dimensionality and then to apply clustering in the projection space.}

Among the MDP techniques, the Force Directed (FD) graph drawing technique \cite{fruchterman1991graph}, the t-distributed Stochastic Neighbor Embedding (t-SNE), \cite{maaten2008visualizing} and the recent Uniform Manifold Approximation and Projection technique (UMAP), \cite{mcinnes2018umap}, have been widely used for dimensionality reduction. They transform high-dimensional data into two-dimensional scatter plot representations while preserving data similarity, and hence possible clusters.

Regarding the clustering techniques that could be used in the resulting 2D space, we can select them based on their lower computational complexity, their ability to find groups with various shapes, and the number of hyper-parameters to tune. Moreover, we are interested in detecting the core clusters that are likely to generate strong stances, rather than noisy sparse clusters with low influence. DBSCAN \cite{ester1996density} and Mean Shift \cite{comaniciu2002mean} are two well-known clustering techniques that satisfy these constraints and further enable the discovery of core clusters and high-density peaks, respectively, with low computational complexity and fewer hyper-parameters to tune.

In this work, we explore combinations of (\emph{i})~relevant input features, namely \textit{retweeted tweets}, \textit{retweeted accounts}, and \textit{hashtags}, (\emph{ii})~dimensionality reduction of these input spaces into two dimensions using FD, t-SNE and UMAP, and (\emph{iii})~clustering thereafter using DBSCAN and Mean Shift, to determine the most efficient pipeline for finding stance clusters (see Figure \ref{fig:pipelineoverview}).

\section{Finding Stance Clusters}

\paragraph{Feature Selection:} Given a tweet dataset that has been pre-filtered using topical words, we take the $n$ most ``engaged'' users who have posted a minimum number of tweets in the dataset.  Given this sample of users, we compute the cosine similarity between each pair of users.  \rev{This similarity can be computed based on a variety of features including (re)tweeting identical tweets, which is what is typically used in label propagation; the hashtags that users use; or the accounts they retweet. Thus, the dimensions of the feature spaces given the different features are the \textit{number of unique tweets} (feature space T), the \textit{number of unique hashtags} (feature space H), or the \textit{number of unique retweeted accounts} (feature space R).}

We computed the cosine similarity using each of these feature spaces independently as well as concatenating all of them together (below, we will refer to this combination as TRH). 
For example, when constructing a user's sparse feature vector using retweeted accounts (Feature space $R$), the elements in the vector would be all 0 except for the retweeted accounts, where it would correspond to their relative frequency, i.e., the number of times the user has retweeted each of them in our dataset divided by the number of times the user has retweeted any of them.  For instance, if the user has retweeted three accounts with frequencies 5, 100, and 895, then the corresponding feature values  
would be 5/1,000, 100/1,000, and 895/1,000, where 1,000 is the sum of the frequencies. 

\begin{figure*}[tbh]
    \centering
    \includegraphics[width=0.81\textwidth]{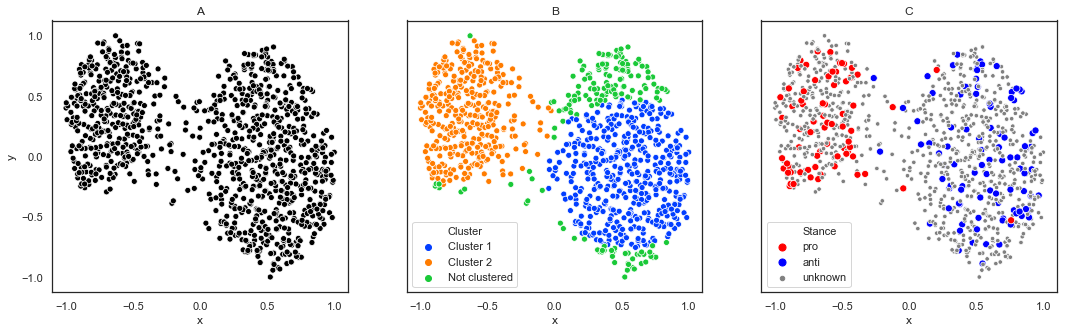}
    \caption{Successful setup:  Plot (A) illustrates how user vectors get embedded by UMAP in two dimensions, Plot (B) presents the clusters Mean Shift produces for them, and Plot (C) shows the users' true labels.}
    \label{fig:UMAP_MeanShift_steps}
\end{figure*}

\begin{figure}[tbh]
    \centering
    \includegraphics[width=.42\textwidth]{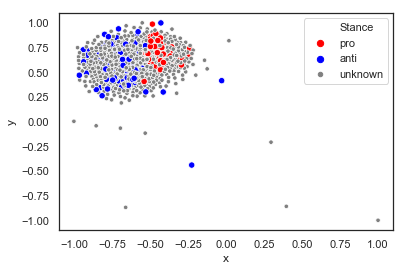}
    \caption{Unsuccessful setup: The user vectors after projection (using t-SNE in this case); the colors show the users' true labels.}
    \label{fig:t-SNE_MeanShift_fail}
\end{figure}

\paragraph{Dimensionality Reduction:} 
We experimented with the following dimensionality reduction techniques based on the aforementioned cosine similarity between users:

\begin{itemize}
\item \textbf{FD} \cite{fruchterman1991graph} minimizes the energy in the representation of the network as a low-dimensional node-link diagram, analogous to a physical system, where edges are springs and nodes bear repulsive charges such that similar nodes are pulled closer together and dissimilar nodes are pushed further apart. In our expеriments, we used the implementation in the NetworkX toolkit.\footnote{\url{http://networkx.github.io/}} 
\item \textbf{t-SNE} \cite{maaten2008visualizing} uses the pre-computed cosine similarity between pairs of users in order to estimate the probability for a user to be the neighbor of another one in the high-dimensional space --- the farther apart they are in terms of cosine similarity, the lower the probability that they are neighbors. A set of points representing the users is located in the low-dimensional space and the same probabilistic matrix is computed based on the relative Euclidean distances in that projection space. The position of the points is updated progressively trying to minimize the Kullback-Leibler divergence between these two probability distributions  \cite{maaten2008visualizing}. In our experiments, we used the scikit-learn\footnote{\url{https://scikit-learn.org}} implementation of t-SNE.
\item \textbf{UMAP} \cite{mcinnes2018umap} is similar to t-SNE, but assumes that the data points are uniformly distributed on a Riemannian connected manifold with a locally constant metric.  A fuzzy topological structure encoded as a weighted K-Nearest Neighbor graph of the data points is used to model that manifold and its uniform metric. The same structure is built in the projection space across the points representing the data, and their position is updated to minimize the divergence between these two structures \cite{mcinnes2018umap}. UMAP is significantly more computationally efficient than t-SNE and tends to emphasize the cluster structure in the projection. We used Leland McInnes's implementation of UMAP.\footnote{\url{http://umap-learn.readthedocs.io/en/latest/}}
\end{itemize}

\paragraph{Clustering:} After projecting the users into a two-dimensional space, we scale user positions in $x$ and $y$ (independently) between $-1$ and $1$ (as shown in the successful plot A in Figure~\ref{fig:UMAP_MeanShift_steps} and in the less successful plot in Figure~\ref{fig:t-SNE_MeanShift_fail}) 
and we proceed to identify cluster cores using the following two clustering methods (see plot B in Figure~\ref{fig:UMAP_MeanShift_steps}):
\begin{itemize}
    \item \textbf{DBSCAN} is a density-based clustering technique which attempts to identify clusters based on preset density \cite{ester1996density}.  It can identify clusters of any shape, but it requires tuning two hyper-parameters related to clustering density: $\epsilon$, which specifies how close the nodes have to be in order to be considered ``reachable'' neighbors, and $m$, which is the minimum number of nodes required to form a core set. Points that are not in a core set nor reachable by any other points are outliers that are not considered as part of the final clusters.  We used the scikit-learn implementation of DBSCAN.
    \item \textbf{Mean Shift} attempts to find peaks of highest density based on a kernel smoothing function \cite{comaniciu2002mean}, typically using a Gaussian kernel. With a  kernel at each point, each point is iteratively shifted to the mean (barycenter) of all the points weighted by its kernel. All points thus converge to the local maximum of the density nearby them. The kernel's bandwidth hyper-parameter determines the number of peaks detected by Mean Shift and all points converging to the same peak are grouped into the same cluster. The bandwidth can be estimated automatically using cross-validation in a probabilistic setting. Orphan peaks where only a few points converge are assumed to be outliers and hence are not clustered.  
    Again, we used the scikit-learn implementation of the algorithm.
\end{itemize}

\paragraph{Labeling Clusters:} Finally, we assume that the users in each cluster would have the same stance with respect to the target topic. As we will show later, we are able to find the most salient retweeted accounts and hashtags for each user cluster using a variant of the valence score \cite{conover2011political}. This score can help when assigning labels to user clusters, based on the most frequent characteristics of the group.

\section{Datasets}
\label{section:dataset}

We used two types of datasets: labeled and unlabeled.
We pre-labeled the former in advance, and then we used it to try different experimental setups and hyper-parameters values. Additionally, we collected fresh unlabeled data on new topics and we applied the best hyper-parameters on this new data.

\subsection{Labeled Datasets}

We used three datasets in different languages:
\begin{enumerate}[wide, labelwidth=!, labelindent=0pt]
    \item \textbf{Kavanaugh dataset (English):} We collected tweets pertaining to the nomination of Judge Kavanaugh to the US Supreme Court in two different time intervals, namely September 28-30, 2018, which were the three days following the congressional hearing concerning the sexual assault allegation against Kavanaugh, and October 6-9, 2018, which included the day the Senate voted to confirm Kavanaugh and the following three days. We collected tweets using the Twarc toolkit,\footnote{\url{https://github.com/edsu/twarc}} where we used both the search and the filtering interfaces to find tweets containing any of the following keywords: \emph{Kavanaugh, Ford, Supreme, judiciary, Blasey, Grassley, Hatch, Graham, Cornyn, Lee, Cruz, Sasse, Flake, Crapo, Tillis, Kennedy, Feinstein, Leahy, Durbin, Whitehouse, Klobuchar, Coons, Blumenthal, Hirono, Booker}, and \emph{Harris}.  These keywords include the judge's name, his main accuser, and the names of the members of the Senate's Judiciary Committee.  In the process, we collected 23 million tweets, authored by 687,194 users.  Initially, we manually labeled the 50 users who posted the highest number of tweets in our dataset. It turned out that 35 of them supported the Kavanaugh's nomination (labeled as \textbf{pro}) and 15 opposed it (labeled as \textbf{anti}).  Next, we used label propagation to automatically label users based on their retweet behavior \cite{darwish2017predicting,kutlu2018devam,magdy2016isisisnotislam}.  The assumption here is that users who retweet a tweet on the target topic are likely to share the same stance as the one expressed in that tweet. Given that many of the tweets in our collection were actually retweets or duplicates, we labeled users who retweeted 15 or more tweets that were authored or retweeted by the \textbf{pro} group with no retweets from the other group as \textbf{pro}.  Similarly, we labeled users who retweeted 6 or more tweets from the \textbf{anti} group and no retweets from the other side as \textbf{anti}. 
    
We chose to increase the minimum number for the \textbf{pro} group as they were over-represented in the initial manually labeled set.  We performed only one label propagation iteration, labeling 48,854 users: 20,098 as \textbf{pro} and 28,756 as \textbf{anti}. Since we do not have gold labels to compare against, we opted to spot-check the results. Thus, we randomly selected 50 automatically labeled accounts (21 \textbf{pro} and 29 \textbf{anti}), and we manually labeled them.  All automatic labels matched the manual labels. As observed, label propagation may require some tuning to work properly, and checks are needed to ensure efficacy. 
    \item \textbf{Trump dataset (English):} We collected 4,152,381 tweets (from 1,129,459 users) about Trump and the 2018 midterm elections from Twitter over a period of three days (Oct. 25-27, 2018) using the following keywords: \emph{Trump, Republican, Republicans, Democrat, Democrats, Democratic, midterm, elections, POTUS (President of the US), SCOTUS (Supreme Court of the US),} and \emph{candidate}.  We automatically labeled 13,731 users based on the hashtags that they used in their account descriptions.  Specifically, we labeled 7,421 users who used the hashtag \texttt{\#MAGA} (Make American Great Again) as \textbf{pro} Trump and 6,310 users who used any of the hashtags \texttt{\#resist}, \texttt{\#resistance}, \texttt{\#impeachTrump}, \texttt{\#theResistance}, or \texttt{\#neverTrump} as \textbf{anti}. We further tried label propagation, but it increased the number of labeled users by 12\% only; thus, we dropped it. In order to check the quality of the automatic labeling, we randomly labeled 50 users, and we found out that for 49 of them, the manual labels matched the automatic ones. 
    \item \textbf{Erdo\u{g}an dataset (Turkish):} We collected a total of 19,856,692 tweets (authored by 3,184,659 users) about Erdo\u{g}an and the June 24, 2018 Turkish elections that cover the period of June 16--23, 2018 (inclusive). Unlike the previous two datasets, which were both in English, this one was in Turkish. We used many election-related terms including political party names, names of popular politicians, and election-related hashtags.  We were interested in users' stance toward Erdo\u{g}an, the incumbent presidential candidate, specifically.  In order to label users with their stance, we made one simplifying assumption, namely that the supporter of a particular political party would be supporting the candidate supported by that party. Thus, we labeled users who use ``AKParti'' (Erdo\u{g}an's party) in their Twitter user name or screen name as \textbf{pro}. Similarly, we labeled users who mentioned other parties with candidates (``CHP'', ``HDP'', or ``IYI'') in their names as \textbf{anti}.  Further, users who used pro-Erdo\u{g}an hashtags, namely \#devam (meaning ``continue'') or \#RTE (``Recep Tayyip Erdo\u{g}an''), or the anti-Erdo\u{g}an hashtag \#tamam (``enough'') in their profile description as \textbf{pro} or \textbf{anti}, respectively. In doing so, we were able to automatically tag 2,684 unique users: 1,836 as \textbf{pro} and 848 as \textbf{anti}. We further performed label propagation where we labeled users who retweeted ten or more tweets that were authored or retweeted by either the \textbf{pro} or the \textbf{anti} groups, and who had no tweets from the other side.  This resulted in 233,971 labeled users of which 112,003 were \textbf{pro} and 121,968 were \textbf{anti}. We manually labeled 50 random users, and we found out that our manual labels agreed with the automatic ones for 49 of them.
\end{enumerate}

\subsection{Unlabeled Datasets}

Next, we collected fresh tweets on several new topics, which are to be used to test our framework with the best settings we could find on the above labeled datasets.  In particular, we collected tweets on six polarizing topics in USA, as shown in Table~\ref{tab:topics}.  The topics include a mixture of long-standing issues such as \emph{immigration} and \emph{gun control}, transient issues such as the \emph{controversial remarks by Representative Ilhan Omar on the Israeli lobby}, and non-political issues such as the \emph{benefits/dangers of vaccines}. We filtered the tweets, keeping only those by users who had indicated the USA as their location, which we determined using a gazetteer that includes variants of USA, e.g.,~\emph{USA}, \emph{US}, \emph{United States}, and \emph{America}, as well as state names along with their abbreviations, e.g.,~\emph{Maryland} and \emph{MD}.

\begin{table}[ht!]
    \centering
    \scriptsize
    \begin{tabular}{p{.9cm}p{4.2cm}p{.8cm}p{0.8cm}}
    \toprule
    Topic & Keywords & Date Range & No. of Tweets \\ 
    \midrule
    Gun control/rights & \#gun, \#guns, \#weapon, \#2a, \#gunviolence, \#secondamendment, \#shooting, \#massshooting, \#gunrights, \#GunReformNow, \#GunControl, \#NRA & Feb 25--Mar 3, 2019 & 1,782,384 \iffalse7,583,311\fi \\ 
    \midrule
    Ilhan Omar remarks on Israel lobby & IlhanOmarIsATrojanHorse, \#IStandWithIlhan, \#ilhan, \#Antisemitism, \#IlhanOmar, \#IlhanMN, \#RemoveIlhanOmar, \#ByeIlhan, \#RashidaTlaib, \#AIPAC, \#EverydayIslamophobia, \#Islamophobia, \#ilhan & Mar 1--9, 2019 & 2,556,871 \\ 
    \midrule
    Illegal immigration & \#border, \#immigration, \#immigrant, \#borderwall, \#migrant, \#migrants, \#illegal, \#aliens & Feb 25--Mar 4, 2019 & 2,341,316 \iffalse7,593,058\fi \\ \hline
    Midterm & midterm, election, elections & Oct 25--27, 2018 & 520,614 \\ 
    \midrule
    Racism \& police brutality & \#blacklivesmatter, \#bluelivesmatter, \#KKK, \#racism, \#racist, \#policebrutality, \#excessiveforce, \#StandYourGround, \#ThinBlueLine & Feb 25--Mar 3, 2019 & 2,564,784 \iffalse8,369,133\fi \\ 
    \midrule
    Vaccination benefits \& dangers & \#antivax, \#vaxxing, \#BigPharma, \#antivaxxers, \#measlesoutbreak, \#Antivacine, \#VaccinesWork, \#vaccine, \#vaccines, \#Antivaccine, \#vaccinestudy, \#antivaxx, \#provaxx, \#VaccinesSaveLives, \#ProVaccine, \#VaxxWoke, \#mykidmychoice & Mar 1--9, 2019 & 301,209 \\ 
    \bottomrule
    \end{tabular}
    \caption{Controversial topics.}
    \label{tab:topics}
\end{table}

\begin{table*}[tbh]
\begin{center}
\scriptsize
    \begin{tabular}{cccccccrc}
    \toprule
Set	&	\# of Users	&	Feature(s)	&	Dim Reduce	&	Peak Detect	&	Avg. Purity	&	Avg. \# of Clusters	&	Avg. Cluster Size	&	Avg. Recall	\\ \hline
\multirow{5}{*}{100k}	&	\multirow{3}{*}{500}	&	R	& FD	& Mean Shift	& \textbf{90.1}	&	2.0 &	100.9 &	40.4 \\
    &        &    R	&	UMAP	&	Mean Shift	&	86.6	&	2.5	&	125.4	&	50.2	\\
	&		&	TRH	&	UMAP	&	Mean Shift	&	85.5	&	2.0	&	145.9	&	\textbf{58.4}	\\ \cline{2-9}
	&	\multirow{2}{*}{1,000}	&	R	&	UMAP	&	Mean Shift	&	\textbf{90.5}	&	2.9	&	196.1	&	39.2	\\
	&		&	TRH	&	UMAP	&	Mean Shift	&	88.3	&	2.3	&	305.8	&	\textbf{61.2}	\\ \hline
\multirow{9}{*}{250k}	&	\multirow{3}{*}{500} &	R & FD & Mean Shift & \textbf{98.7}	&	2.5	 & 171.3 & 68.6 \\
    &       &	R	&	UMAP	&	Mean Shift	&	98.5	&	2.1	&	179.9	&	\textbf{72.0}	\\
	&		&	TRH	&	UMAP	&	Mean Shift	&	94.4	&	2.3	&	165.3	&	66.2	\\ \cline{2-9}
	&	\multirow{3}{*}{1,000} &	R & FD & Mean Shift & 	\textbf{99.1}	&	2.3 & 	353.5 &	70.6 \\
	&       &	R	&	UMAP	&	Mean Shift	&	98.8	&	2.1	&	359.2	&	\textbf{71.8}	\\
	&		&	TRH	&	UMAP	&	Mean Shift	&	97.9	&	2.5	&	355.5	&	71.2	\\ \cline{2-9}
	&	\multirow{3}{*}{5,000}  &	R & FD & Mean Shift & \textbf{98.8}	&	2.1 &	1,264.3	& 50.6 \\
	&       &	R	&	UMAP	&	Mean Shift	&	98.6	&	2.4	&	1,322.2	&	52.8	\\
	&		&	TRH	&	UMAP	&	Mean Shift	&	97.9	&	2.7	&	1,872.4	&	\textbf{74.8}	\\ \hline
\multirow{23}{*}{1M}	&	\multirow{6}{*}{500}	&	R	&	FD	&	Mean Shift	&	\textbf{99.0}	&	2.6 & 180.4 & 72.2	\\
	&		&	R	&	t-SNE	&	Mean Shift	&	94.9	&	2.1	&	165.1	&	66.0	\\
	&		&	R	&	UMAP	&	Mean Shift	&	97.5	&	2.6	&	179.8	&	72.0	\\
	&		&	T	&	UMAP	&	Mean Shift	&	98.0	&	2.0	&	162.3	&	65.0	\\
	&		&	TRH	&	t-SNE	&	Mean Shift	&	91.7	&	2.3	&	171.3	&	68.6	\\
	&		&	TRH	&	UMAP	&	Mean Shift	&	98.9	&	2.3	&	186.5	&	\textbf{74.6}	\\ \cline{2-9}
	&	\multirow{10}{*}{1,000}	&	R	&	FD	&	Mean Shift	&	\textbf{99.4}	&	2.1 & 366.7 & 73.4	\\
	&		&	R	&	t-SNE	&	Mean Shift	&	94.6	&	2.0	&	309.9	&	62.0	\\
	&		&	R	&	UMAP	&	DBSCAN	&	84.4	&	2.2	&	403.1	&	80.6	\\
	&		&	R	&	UMAP	&	Mean Shift	&	98.9	&	2.7	&	369.5	&	73.8	\\
	&		&	T	&	t-SNE	&	Mean Shift	&	92.7	&	2.0	&	307.7	&	61.6	\\
	&		&	T	&	UMAP	&	Mean Shift	&	98.6	&	2.0	&	349.8	&	70.0	\\
	&		&	TRH	&	FD	&	Mean Shift	&	95.7	&	2.1	&	326.3	&	65.2	\\
	&		&	TRH	&	t-SNE	&	Mean Shift	&	96.0	&	2.1	&	348.1	&	69.6	\\
	&		&	TRH	&	UMAP	&	DBSCAN	&	81.7	&	2.0	&	415.1	&	\textbf{83.0}	\\
	&		&	TRH	&	UMAP	&	Mean Shift	&	98.7	&	2.7	&	366.8	&	73.4	\\ \cline{2-9}
	&	\multirow{7}{*}{5,000}	&	R	&	FD	&	Mean Shift	&	\textbf{99.6}	&	2.3 & 1,971.5 & 78.8\\
	&		&	R	&	UMAP	&	Mean Shift	&	99.3	&	2.5	&	1,965.2	&	78.6	\\
	&		&	T	&	t-SNE	&	Mean Shift	&	97.8	&	2.0	&	1,795.0	&	71.8	\\
	&		&	T	&	UMAP	&	Mean Shift	&	99.2	&	2.1	&	1,869.3	&	74.8	\\
	&		&	TRH	&	FD	&	Mean Shift	&	99.1	&	2.0	&	1,838.8	&	73.6	\\
	&		&	TRH	&	UMAP	&	DBSCAN	&	93.2	&	2.2	&	2,180.6	&	\textbf{87.2}	\\
	&		&	TRH	&	UMAP	&	Mean Shift	&	99.4	&	2.3	&	1,980.7	&	79.2	\\
	\bottomrule
    \end{tabular}
    \caption{Results for combinations that meet the success criteria: at least 2 clusters, average label purity of at least 80\% across all clusters, and labels assigned to at least 30\% of the available users.  The table shows the average purity, the average number of clusters, the average number of users who were automatically tagged, and the average proportion of users who were tagged (Recall) across the 15 tweet subsets.}
    \label{table:successfullSetupsResults}
\end{center}
\end{table*}

\begin{table}[tbh]
\begin{center}
\scriptsize
  \begin{tabular}{C{1.4cm}C{1.2cm}C{0.85cm}C{0.85cm}C{0.85cm}R{0.85cm}}
  \toprule
    Dim-Reduce param & Peak-Detect param & Avg. Purity & Avg. \# of Clusters & Avg. Cluster Size  & Avg. Run Time (s) \\ \hline
    \multicolumn{6}{c}{FD+Mean Shift} \\ \hline
    -            & bin=False & 99.0 & 2.2 & 356.8 & 226   \\
    -            & bin=True & \textbf{99.2} & 2.1 & 356.0   & \textbf{191}   \\ \hline
    \multicolumn{6}{c}{UMAP+Mean Shift} \\ \hline
    neighbors=15 &  bin=False & \textbf{98.6} & 2.0 & 354.3 & 148   \\
    neighbors=15 &  bin=True & 98.4 & 2.0 & 348.9 & \textbf{78}    \\
    neighbors=5 &  bin=True &\textbf{ 98.6} & 2.0 & 358.2 & 114   \\
    neighbors=10 &  bin=True & \textbf{98.6} & 2.0 & 353.2 & 129   \\
    neighbors=20 &  bin=True & 98.4 & 2.0 & 348.7 & 159   \\
    neighbors=50 &  bin=True & 98.4 & 2.0 & 353.7 & 159   \\
    \bottomrule
  \end{tabular}
  \caption{Sensitivity of FD+Mean Shift and UMAP+Mean Shift to hyper-parameter variations and random initialization. Experiments on 250k datasets, top 1,000 users, and using R to compute similarity. For UMAP, we tuned n\_neighbors (default=15), and for Mean Shift we ran with and without bin\_seeding (default=True).}
  \label{table:moreExperimentsSensitivity}
\end{center}
\end{table}

\section{Experiments and Evaluation}

\subsection{Experimental Setup} 

\rev{We randomly sampled tweets from each of the datasets to create datasets of sizes 50k, 100k, 250k, and 1M. For each subset size (e.g., 50k), we created 5 sub-samples of the three datasets to create 15 tweet subsets, on each of which we ran a total of 72 experiments with varying setups:}
\begin{itemize}
    \item The dimensionality reduction technique: FD, t-SNE, or UMAP. \rev{FD needs no hyper-parameter tuning.} We used the default hyper-parameters for t-SNE and UMAP \rev{(we change these defaults below)}: for t-SNE, we used \textit{perplexity} $\rho=30.0$ and \textit{early exaggeration} $ee=12.0$, while for UMAP, we used \textit{n\_neighbors}=15 and \textit{min\_distance}=0.1.
    \item The peak detection/clustering algorithm: DBSCAN or Mean Shift. We used the default hyper-parameters for DBSCAN, namely $\epsilon$=0.5 and $m$=5. For Mean Shift, the bandwidth hyper-parameter was estimated automatically as the threshold for outliers. 
    \item The number of top users to cluster: 500, 1,000, or 5,000.  Clustering a smaller number of users requires less computation. We only considered users with a minimum of 5 interactions, e.g.,~5 retweeted tweets.
    \item The features used to compute the cosine similarity, namely Retweets (R), Hashtags (H), full Tweets (T), or all of them together (TRH).
\end{itemize}

\subsection{Evaluation Results} 

We considered a configuration as effective\rev, i.e.,~successful, if it yielded a few mostly highly pure clusters with a relatively low number of outliers, namely with an average label \textit{purity of at least 80\%} across all clusters and where labels are \textit{assigned to at least 30\%} of the users that were available for clustering. Since polarizing topics typically have two main sides, the number of generated clusters would ideally be 2 (or perhaps 3) clusters.

Table \ref{table:successfullSetupsResults} lists all results for experimental configurations that meet our success criteria. Aside from the parameters of the experiments, we further report on average cluster purity, average number of clusters, average cluster size, and average recall, which is the number of users in the same cluster.  A few observations can be readily gleaned from the results, namely:
\begin{itemize}
    \item No setup involving 50k subsets met our criteria, but many larger setups did.  Purity increased between 8.3-11.9\% on identical setups when moving from 100k to 250k, while the improvement in purity was mixed when using the 1M tweet subsets compared to using 250k.
    \item All setups meeting our criteria when using the 100k and 250k subsets involved using retweets as a feature (R or TRH), FD or UMAP for dimensionality reduction, and Mean Shift for peak detection. Some other configurations met our criteria only when using subsets of size 1M. 
    \item Using retweets (R) 
    to compute similarity yielded the highest purity when using 1M tweets, 5,000 users, FD, and Mean Shift with purity of 99.6\%. 
    Note that this setup is quite computationally expensive.
    \item Using hashtags (H) alone to compute similarity failed to meet our criteria in all setups.
\end{itemize}

As mentioned earlier, reducing the size of the tweet sets and the number of users we cluster would lead to greater computational efficiency. Thus, based on the results in Table~\ref{table:successfullSetupsResults}, we focused on the setup with 250k tweets, 1,000 users, retweets (R) as feature, FD or UMAP for clustering, and Mean Shift for peak detection. This setup yielded high purity (99.1\% for FD and 98.8\% UMAP) that is slightly lower than our best results (99.6\%: 1M tweets, R as feature, FD, and Mean Shift) while being relatively more computationally efficient  than the overall best setup.  

We achieved the best purity with two clusters on average when the dimensionality reduction method used the FD algorithm and the clustering method was Mean Shift.
However, as shown in Table~\ref{table:moreExperimentsSensitivity}, UMAP with Mean Shift yielded similar purity and cluster counts, while being more than twice as fast as FD with Mean Shift.

\rev{
\subsection{The Role of Dimensionality Reduction}

We also tried to use Mean Shift to cluster users directly without performing dimensionality reduction, but we found that Mean Shift alone was never able to produce clusters that meet our success criteria, despite automatic and manual hyper-parameter tuning. Specifically, we experimented on the subsets of size 250k. Mean Shift failed to produce more than one cluster with the cluster subsuming more than 95\% of the users.
} 

\rev{

\subsection{Comparison to Supervised Classification}
We compared our results to using supervised classification of users.  For each of the 250k sampled subsets for each of the three labeled datasets, we retained users for which we have stance labels and we randomly selected 100 users for training and the remaining users for testing.  We used the retweeted accounts for each user as features.  We used two different classifiers, namely SVM$^{light}$, which is a support vector machine (SVM) classifier, and fastText, which is a deep learning classifier \cite{joulin2016bag}. 

\begin{table}[h!]
\begin{center}
\scriptsize
\begin{tabular}{cccc}
    \toprule
     Classifier & Precision & Recall & F1 \\ \midrule
     SVM$^{light}$ & 86.0\% & 95.3\% & 90.4\% \\
     fastText & 64.2\% & 64.2\% & 64.2\% \\
     \bottomrule
\end{tabular}
\end{center}
\caption{Results for supervised classification.}
\label{table:supervised}
\end{table}

The evaluation results are shown in Table~\ref{table:supervised}. To measure classification effectiveness, we used precision, recall, and F1 measure. We can see that SVM$^{light}$ outperforms fastText by a large margin. The average cluster purity and the average recall in our results for our unsupervised method (see Table~\ref{table:successfullSetupsResults}) are analogous to the precision and the recall in the supervised classification setup, respectively. 

Comparing to our unsupervised method (250k tweet subset, 1,000 users, R as feature, UMAP, and Mean Shift), we can see that our method performs better than the SVM-based classification in terms of precision (99.1\% cluster purity compared to 86.0\%), but has lower recall (70.6\% compared to 95.3\%). 
However, given that our unsupervised method is intended to generate a core set of labeled users with very high precision, which can be used to train a subsequent classifier, e.g., a tweet-based classifier, without the need for manual labeling, precision is arguably more important than recall. 
}

\begin{table}[tbh!]
\begin{center}
\scriptsize
  \begin{tabular}{cC{0.8cm}cC{0.9cm}C{0.9cm}R{0.9cm}}
    \toprule
  Dim-Reduce & Peak-Detect & Avg. Purity    & Avg. \# of Clusters & Avg. Cluster Size  & Run Time (s) \\ \hline
    \multicolumn{6}{c}{t-SNE+Mean Shift (bin\_seeding=True)} \\ \hline
    $\rho$=30/ee=8                       & -                    & 69.7        & 1.6           & 256.0 & 290   \\
    $\rho$=30/ee=12                      & -                    & 69.5        & 1.6           & 260.6 & 286   \\
    $\rho$=30/ee=50                      & -                    & 69.6        & 1.8           & 266.6 & 301   \\
    \textbf{$\rho$=5/ee=8}               & -                    & \textbf{98.0}        & \textbf{2.0}             & 358.0 & 190   \\
    \textbf{$\rho$=5/ee=12 }             & -                    & \textbf{98.2 }       & \textbf{2.0}             & 359.1 & 193   \\
    \textbf{$\rho$=5/ee=50  }            & -                    & \textbf{98.4}        & \textbf{2.0}             & 360.0 & 192   \\
    \rev{$\rho$=5/dim=3}             & -                    & \rev{60.2}        & \rev{1.0}             & \rev{238.2} & \rev{589}   \\\hline
    \multicolumn{6}{c}{UMAP (n\_neighbors=15)+DBSCAN} \\ \hline
    -                            & $\epsilon$=0.50                       & 70.4        & 1.3           & 410.5 & 74    \\
    -                            & \textbf{$\epsilon$=0.10}              & \textbf{95.9$\pm$1.7}   & \textbf{2.3$\pm$0.1}     & 408.9 & 73    \\
    -                            & \textbf{$\epsilon$=0.05}             & \textbf{98.9}        & \textbf{16.8}          & 341.1 & 78    \\\hline
    \multicolumn{6}{c}{t-SNE+DBSCAN} \\ \hline
    $\rho$=30/ee=8              & $\epsilon$=0.50                       & 59.5        & 1             & 409.9 & 195   \\
    $\rho$=30/ee=12             & $\epsilon$=0.50                       & 59.5        & 1             & 409.9 & 192   \\
    $\rho$=30/ee=50             & $\epsilon$=0.50                       & 59.5        & 1             & 409.7 & 201   \\
    $\rho$=30/ee=8              & $\epsilon$=0.10              & 59.2        & 1             & 397.9 & 184   \\
    $\rho$=30/ee=12             & $\epsilon$=0.10              & 59.3        & 1             & 397.3 & 193   \\
    $\rho$=30/ee=50             & $\epsilon$=0.10              & 59.2        & 1             & 397.6 & 195   \\
    $\rho$=5/ee=8               & $\epsilon$=0.50               & 59.5        & 1             & 410.0   & 135   \\
    $\rho$=5/ee=12              & $\epsilon$=0.50               & 59.5        & 1             & 410.0   & 135   \\
    $\rho$=5/ee=50              & $\epsilon$=0.50               & 59.5        & 1             & 410.0   & 148   \\
    $\rho$=5/ee=8               & $\epsilon$=0.10              & 71.8$\pm$1.5 & 1.6$\pm$0.1     & 407.4 & 140   \\
    $\rho$=5/ee=12              & $\epsilon$=0.10              & 74.0$\pm$2.2 & 1.7$\pm$0.1     & 407.0 & 131   \\
    $\rho$=5/ee=50              & $\epsilon$=0.10              & 75.5$\pm$2.1 & 1.6$\pm$0.1     & 407.0 & 139   \\\hline
    \multicolumn{6}{c}{FD+DBSCAN} \\ \hline
    -            & $\epsilon$=0.50                       & 59.5        & 1             & 410.4 & 179   \\
    -            & $\epsilon$=0.10             & 70          & 1.3           & 399.1 & 177   \\
    -            & $\epsilon$=0.05         & 78.1        & 1.7           & 372.5 & 178   \\
    \bottomrule
  \end{tabular}
  \caption{Sensitivity of t-SNE and DBSCAN to changes in hyper-parameter values and to random initialization. The experiments ran on the 250k datasets, 1,000 most engaged users, and using R to compute similarity. For t-SNE, we experimented with perplexity $\rho\in\{5,30*\}$, early exaggeration $ee\in\{8, 12*, 50\}$\rev{, and number of dimensions of output $dim\in\{2*, 3\}$}.  For DBSCAN, we varied epsilon $\epsilon\in\{0.05, 0.50*\}$. $*$ means default value. Only the numbers with stdev$>$0.0 over multiple runs show stdev values after them. Entries meeting our success criteria are bolded.
  }
  \label{table:moreExperimentsTuning}
\end{center}
\end{table}

\begin{table*}[tbh!]
\begin{center}
\scriptsize
    \begin{tabular}{lp{4cm}c||lp{4cm}c}
    \toprule
\multicolumn{6}{c}{\bf Kavanaugh Dataset} \\
\multicolumn{3}{c||}{Cluster 0 (Left-leaning)}					&	\multicolumn{3}{c}{Cluster 1 (Right-leaning)}					\\ \hline
RT	&	Description	&	score	&	RT	&	Description	&	score	\\ \hline
@kylegriffin1	&	Producer. MSNBC's @TheLastWord. 	&	55.0	&	@mitchellvii	&	(pro-Trump) Host of YourVoice™ America	&	52.5	\\
@krassenstein	&	Outspoken critic of Donald Trump - Editor at http://HillReporter.com	&	34.0	&	@FoxNews	&	(right leaning media)	&	48.0	\\
@Lawrence	&	thelastword.msnbc.com 	&	29.0	&	@realDonaldTrump	&	45th President of the United States	&	48.0	\\
@KamalaHarris	&	(Dem) U.S. Senator for California.	&	29.0	&	@Thomas1774Paine	&	TruePundit.com 	&	47.0	\\
@MichaelAvenatti	&	(anti-Trump) Attorney, Advocate, Fighter for Good.	&	26.0	&	@dbongino	&	Host of Dan Bongino Podcast. Own the Libs.	&	44.5	\\ \hline 
Hashtag	&	Description	&	score	&	Hashtag	&	Description	&	score	\\ \hline
StopKavanaugh	&	-	&	5.0	&	ConfirmKavanaugh	&	-	&	19.0	\\
SNL	&	Saturday Night Live (ran a skit mocking Kavanaugh)	&	4.0	&	winning	&	pro-Trump	&	12.0	\\
P2	& progressives on social media		&	3.0	&	Qanon	&	alleged insider/conspiracy theorist (pro-Trump)	&	11.0	\\
DevilsTriangle	&	sexual/drinking game	&	3.0	&	WalkAway	&	walk away from liberalism/Dem party	&	9.0	\\
MSNBC	&	left-leaning media	&	3.0	&	KavanaughConfirmation	&		&	8.0	\\ 
\bottomrule
\toprule
\multicolumn{6}{c}{\bf Trump Dataset} \\
\multicolumn{3}{c||}{Cluster 0 (Left-leaning)}					&	\multicolumn{3}{c}{Cluster 1 (Right-leaning)}					\\ \hline
RT	&	Description	&	score	&	RT	&	Description	&	score	\\ \hline
@TeaPainUSA	&	Faithful Foot Soldier of the \#Resistance	&	98.5	&	@realDonaldTrump	&	45th President of the United States	&	95.4	\\
@PalmerReport	&	Palmer Report: Followed by Obama. Blocked by Donald Trump Jr	&	69.8	&	@DonaldJTrumpJr	&	EVP of Development \& Acquisitions The @Trump Org	&	72.4	\\
@kylegriffin1	&	Producer. MSNBC's @TheLastWord. 	&	66.5	&	@mitchellvii	&	(pro-Trump) Host of YourVoice™ America	&	47.9	\\
@maddow	&	rachel.msnbc.com 	&	39.5	&	@ScottPresler	&	 spent 2 years to defeat Hillary. I'm voting for Trump	&	33.0	\\
@tribelaw	&	(anti-Trump Harvard faculty)	&	32.0	&	@JackPosobiec	&	OANN Host. Christian. Conservative. 	&	32.5	\\ \hline
Hashtag	&	Description	&	score	&	Hashtag	&	Description	&	score	\\ \hline
VoteBlue	&	Vote Dem	&	12	&	Fakenews	&		&	18.5	\\
VoteBlueToSaveAmerica	&	Vote Dem	&	11	&	Democrats	&	-	&	15.5	\\
AMJoy	&	program on MSNBC	&	5	&	LDtPoll	&	Lou Dobbs (Fox news) poll	&	12.0	\\
TakeItBack	&	Democratic sloagan	&	4	&	msm	&	main stream media	&	11.0	\\
Hitler	&	controvercy over the term "nationalist"	&	3	&	FakeBombGate	&	claiming bombing is fake	&	11.0	\\ \bottomrule
\toprule
\multicolumn{6}{c}{\bf Erdo\u{g}an Dataset} \\
\multicolumn{3}{c||}{Cluster 0 (anti-Erdo\u{g}an)}					&	\multicolumn{3}{c}{Cluster 1 (pro-Erdo\u{g}an)}					\\ \hline
RT	&	Description	&	score	&	RT	&	Description	&	score	\\ \hline
@vekilince	&	(Muhammem Inci -- presidential candidate)	&	149.6	&	@06melihgokcek	&	(Ibrahim Melih Gokcek -- ex. Governer of Ankara)	&	64.9	\\
@cumhuriyetgzt	&	(Cumhuriyet newspaper)	&	104.0	&	@GizliArsivTR	&	(anti-Feto/PKK account)	&	54.0	\\
@gazetesozcu	&	(Sozcu newspaper)	&	82.5	&	@UstAkilOyunlari	&	(Pro-Erdo\u{g}an conspiracy theorist)	&	49.7	\\
@kacsaatoldunet	&	(popular anti-Erdo\u{g}an account) 	&	80.0	&	@medyaadami	&	(Freelance journalist)	&	42.0	\\
@tgmcelebi	&	(Mehmet Ali Celebi -- leading CHP member)	&	65.8	&	@Malazgirt\_Ruhu	&		&	37.0	\\ \hline
Hashtag	&	Description	&	score	&	Hashtag	&	Description	&	score	\\ \hline
tamam	&	enough (anti-Erdo\u{g}an)	&	49.0	&	VakitT{\"u}rkiyeVakti	&	AKP slogan “It is Turkey time”	&	42.7	\\
Muharrem{\.I}ncee	&	Muharrem {\.I}nce -- presidential candidate	&	43.5	&	i̇yikiErdoğanVar	& Great that Erdo\u{g}an is around	&	20.0	\\
demirta{\c s}	&	Selahattin Demirta{\c s} -- presidential candidate	&	12.0	&	tatanka	& Inci's book of poetry		&	19.0	\\
K{\i}l{\i}{\c c}daro\u{g}luNeS{\" o}yledi	&	“what did K{\i}l{\i}{\c c}daro\u{g}lu (CHP party leader) say”	&	11.0	&	Haz{\i}r{\i}zT{\"u}rkiye	&	Turkey: We're Ready	(AKP slogan) &	17.7	\\
mersin	&	place for Inci rally	&	11.0	&	katilHDPKK	&	Killer PKK (Kurdish group)	&	17.0	\\
    \bottomrule
    \end{tabular}
    \caption{Salient retweeted accounts (top 5) and hashtags (top 5) for the two largest clusters for 250k sampled subsets from the Kavanaugh, Trump, and Erdo\u{g}an datasets to qualitatively show the efficacy of our method.  When describing the Twitter accounts, we tried to use the text in the account descriptions as much as possible, with our words put in parentheses.
    }
    \label{table:ClusterLabels}
\end{center}
\end{table*}

\rev{
\subsection{Experiments on New Unlabeled Data}

Next, we experimented with new unlabeled data, as described above. In particular, we used the tweets from the six topics shown in Table~\ref{tab:topics}.  For all experiments, we used UMAP and Mean Shift for dimensionality reduction and clustering, respectively, and we clustered the top 1,000 users using retweets in order to compute similarity. To estimate the cluster purity, we randomly selected 25 users from the largest two clusters for each topic.  A human annotator with expertise in US politics manually and independently tagged the users with their stances on the target topics (e.g.,~pro-gun control/pro-gun rights; pro-DNC/pro-GOP for midterm elections).  

Given the manual labels, we found that the average cluster purity was 98.0\% with an average recall of 86.5\%.  As can be seen, the results are consistent with the previous experiments on the labeled sampled subsets.}

\section{Analysis: Refining in Search of Robustness}

Thus far, we used the default hyper-parameters for all dimensionality reduction and peak detection algorithms.  In the following, we conduct two additional sets of experiments on the 250k dataset, using retweets (R) as features, and the 1,000 most active users.  In the first experiment, we want to ascertain the robustness of our most successful techniques to changes in hyper-parameters and to initialization.  In contrast, in the second experiment, we aim to determine whether we can get other setups to work by tuning their hyper-parameters. 

\subsection{Testing the Sensitivity of the Successful Setups} 

Our successful setups involved using FD or UMAP for dimensionality reduction and Mean Shift for peak detection. \rev{Varying the number of dimensions for dimensionality reduction for both FD and UMAP did not change the results. Thus, we fixed this number to 2 and we continued testing the sensitivity of other hyper-parameters.} FD does not have any tunable hyper-parameters aside from the dimensions of the lower dimensional space, which we set to 2, and the number of iterations, which is by default set to 50.  For UMAP, we varied the number of neighbors ($n\_neighbors$), trying 5, 10, 15, 20, and 50, where 15 was the default.  Mean Shift has two hyper-parameters, namely the bandwidth and a threshold for detecting orphan points, which are automatically estimated by the scikit-learn implementation. 

\noindent As for the rest, we have the option to use bin seeding or not, and whether to cluster all points.
Bin seeding involves dividing the space into buckets that correspond in size to the bandwidth to bin the points therein. We experimented with using bin seeding or not, and we chose not to cluster all points but to ignore orphans.

Lastly, since FD and UMAP are not deterministic and might be affected by random initialization, we ran all FD+Mean Shift and UMAP+Mean Shift setups five times to assess the stability of the results. Ideally, we should get very similar values for purity, the same number of clusters, and very similar number of clustered users. 
Table~\ref{table:moreExperimentsSensitivity} reports the results when varying the hyper-parameters for UMAP and Mean Shift.  We can see that there was very little effect on purity, cluster count, and cluster sizes.  Moreover, running the experimental setups five times always yielded identical results.
Concerning timing information, using binning (\textit{bin\_seeding}=True) led to significant speedup.  Also, increasing the number of neighbors generally increased the running time with no significant change in purity. Lastly, UMAP+Mean Shift was much faster than FD+Mean Shift. \rev{Based on these experiments, we can see that FD, UMAP, and Mean Shift were robust to changes in hyper-parameters; using default parameters yielded nearly the best results.}

\subsection{Tuning the Unsuccessful Setups} 

Our unsuccessful setups involved the use of t-SNE for dimensionality reduction and/or DBSCAN for peak detection. We wanted to see whether their failure was due to improper hyper-parameter tuning, and if so, how sensitive they are to hyper-parameter tuning. t-SNE has two main hyper-parameters, namely \textit{perplexity}, which is related to the size of the neighborhood, and \textit{early exaggeration}, which dictates how far apart the clusters would be placed.  DBSCAN has two main hyper-parameters, namely \textit{minimum neighborhood size} ($m$) and \textit{epsilon} ($\epsilon$), which is the minimum distance between the points in a neighborhood.  Due to the relatively large number of points that we are clustering, $\epsilon$ is the important hyper-parameter to tune, and we experimented with $\epsilon$ equal to 0.50 (default), 0.10, and 0.05. Table~\ref{table:moreExperimentsTuning} reports on the results of hyper-parameter tuning.  As can be seen, no combination of t-SNE or FD with DBSCAN met our minimum criteria (purity $\geq$ 0.8, no. of clusters $\geq$ 2).  t-SNE worked with Mean Shift when \textit{perplexity} ($\rho$) was lowered from 30 (default) to 5. Also, t-SNE turned out to be insensitive to its \textit{early exaggeration} ($ee$) hyper-parameter. \rev{We also experimented by raising the dimensionality of the output of t-SNE, which significantly lowered the purity as well as increased the running time.} UMAP worked with DBSCAN when $\epsilon$ was set to 0.1.  Higher values of $\epsilon$ yielded low purity and too few clusters, while lower values of $\epsilon$ yielded high purity but too many clusters.  Thus, DBSCAN is sensitive to hyper-parameter selection. Further, when we ran the UMAP+DBSCAN setup multiple times, the results varied considerably, which is also highly undesirable.  

Based on these experiments, we can conclude that using FD or UMAP for dimensionality reduction in combination with Mean Shift yields the best results in terms of cluster purity and recall with robustness to hyper-parameter setting. 

Lastly, we found that the execution times of Mean Shift and of DBSCAN were comparable, and UMAP ran significantly faster than FD. 

Therefore, we recommend the following setup for automatic stance detection: UMAP + Mean Shift with the default settings as set in scikit-learn.

\subsection{Labeling the Clusters}  

We wanted to elucidate the cluster outputs by identifying the most salient retweeted accounts and hashtags in each of the clusters.  Retweeted accounts and hashtags can help tag the resulting clusters.  To compute a salience score for each element (retweeted account or hashtag), we initially computed a modified version of the valence score \cite{conover2011political} at accommodates for having more than two clusters.  The valence score ranges in value between $-1$ and $1$, and it is computed for an element $e$ in cluster $A$ as follows:
\begin{equation}
    V(e) = 2 \frac{
    \frac{tf_{A}}{total_{A}}}
    {\frac{tf_{A}}{total_{A}} + \frac{tf_{\lnot A}}{total_{\lnot A}}} -1
    \label{eq:valence}
\end{equation}
where $tf$ is the \textit{frequency} of the element in either cluster $A$ or not in cluster $A$ ($\lnot A$) and $total$ is the sum of all $tf$s for either $A$ or $\lnot A$. We only considered terms that yielded a valence score $V(e) \geq 0.8$.  Next, we computed the score of each element as its frequency in cluster $A$ multiplied by its valence score as $score(e) = tf(e)_A \bullet V(e)$.  Table~\ref{table:ClusterLabels} shows the top 5 retweeted accounts and the top 5 hashtags for 250k sampled sets for all three datasets. 
As the entries and their descriptions in the table show, the salient retweeted accounts and hashtags clearly illustrate the stance of the users in these clusters, and hence can be readily used to assign labels to the clusters. For example, the top retweeted accounts and hashtags for the two main clusters for the Kavanaugh and Trump datasets clearly indicate right- and left-leaning clusters.  A similar picture is seen for the Erdo\u{g}an dataset clusters.

\section{Conclusion and Future Work}

We have presented an effective unsupervised method for identifying clusters of Twitter users who have similar stances with respect to controversial topics. Our method uses dimensionality reduction followed by peak detection/clustering. It overcomes key shortcomings of pre-exiting stance detection methods, which rely on supervised or semi-supervised classification, with the need for manual labeling of many users, which requires both topic expertise and time, and are sensitive to skews in the distribution of the classes in the dataset.  

For dimensionality reduction, we experimented with three different methods, namely Fruchterman-Reingold force-directed algorithm, t-SNE, and UMAP. Dimensionality reduction has several desirable effects such as bringing together similar items while pushing dissimilar items further apart in a lower dimensional space, visualizing data in two dimensions, which enables an observer to ascertain how separable users stances are, and enabling the effective use of downstream clustering. For clustering, we experimented with DBSCAN and Mean Shift, both of which are suited for identifying clusters of arbitrary shapes and are able to identify cluster cores while ignoring outliers.  We conducted a large set of experiments using different features to compute the similarity between users on datasets of different sizes with varying topics and languages that were independently labeled with a combination of manual and automatic techniques. 

\noindent Our most accurate setups use retweeted accounts as features, either the Fruchterman-Reingold force-directed algorithm or UMAP for dimensionality reduction, and Mean Shift for clustering, with UMAP being significantly faster than Fruchterman-Reingold. These setups were able to identify groups of users corresponding to the predominant stances on controversial topics with more than 98\% purity based on our benchmark data. We were able to achieve these results by working with the most active 500 or 1,000 users in tweet sets containing 250k tweets. We have also shown the robustness of our best setups to variations in the algorithm hyper-parameters and with respect to random initialization. 

In future work, we want to use our stance detection technique to profile popularly retweeted Twitter users, cited websites, and shared media by ascertaining their valence scores across a variety of polarizing topics.  

\bibliographystyle{aaai}
\bibliography{references}

\end{document}